\documentclass[12pt]{iopart}
\usepackage{iopams}
\usepackage{authblk}
\usepackage{graphicx}
\usepackage{bm}
\usepackage{amssymb}
\usepackage{color}
\usepackage{hyperref}
\usepackage{amsmath}
\usepackage{amsthm}
\usepackage{mathrsfs}
\usepackage{amsfonts}
\usepackage{cite}
\usepackage{booktabs}

\title{A formalism of Gravitation based on a Physical Field Strength}

\author{L. Horoto\textsuperscript{1}\thanks{22408916@sun.ac.za}, F.G. Scholtz\textsuperscript{1} }
\affil{\textsuperscript{1}Department of Physics, Stellenbosch University, Stellenbosch, South Africa}

\begin{document}
\maketitle
\begin{abstract}
We propose a reformulation of gravitation in which the gravitational interaction is treated as a genuine force rather than an inertial effect arising from spacetime geometry. Within this framework, the difference between the affine connection and a flat reference connection defines a tensor $\mathrm{K}^\mu{}_{\alpha\beta}$, identified as the gravitational field strength. This object cannot be eliminated by coordinate transformations, demonstrating that gravity possesses true physical degrees of freedom. The formalism introduces vector fields $\xi_a{}^\mu$ that extend the notion of infinitesimal translations to curved spacetime and naturally yield a gauge-invariant field strength $\mathfrak{F}^{\xi a}{}_{\mu\nu}$. The dynamics of the gravitational field are governed by a Lagrangian of Yang--Mills type with an additional scalar degree of freedom $\phi^{2}$, corresponding to the Newtonian potential. In the limit of vanishing gravitational coupling $\mathfrak{g}\to0$, the theory reduces to General Relativity, while for nonzero $\mathfrak{g}$ it constitutes an $\mathrm{SU(2)\times U(1)}$ gauge theory of gravity. The framework provides a unified description in which dark energy emerges as the self-interaction energy of the $\phi$ field, and dark-matter-like effects arise from the extended gravitational degrees of freedom. This formulation offers a consistent bridge between classical and quantum descriptions of gravity and clarifies the conceptual foundations of the gravitational interaction.
\end{abstract}

\section{Introduction}

General Relativity (GR) provides a geometric description of gravitation in which free-fall motion is interpreted as geodesic motion in a curved spacetime. Although this geometric formulation has achieved unparalleled empirical success, it renders gravity conceptually distinct from the other fundamental interactions, which are described by gauge fields acting on a (fixed or weakly dynamical) background. This disparity has motivated long-standing efforts to recast gravity in gauge-theoretic language, from the early work of Utiyama and Kibble to modern gauge–gravity dualities. \cite{Einstein1915,YangMills1954,Utiyama1956,Kibble1961,Hehl1976,Mansouri1976}.

With reference to table 1, we compare notions of inertial force and gravitational force as defined in GR when one attempts to view gravity as a force.  From table 1 it should be clear that the former arises from a change of coordinates in flat spacetime, while the latter  arises from not only a change of coordinates but also some effects pertinent to non-trivial geometry. As such, while they are both not physical in the sense that they can be transformed away by an appropriate choice of frame of reference, they are not the same(see page 243 of \cite{coulson1954classical} for a detailed discussion on the difference between gravitational effects and inertial effects). The physical content of gravity is clearly what is left after subtracting the inertial effects. We refer to this as the gravitational aspect of motion.  We note that although the metric in the latter case can be made locally flat, as dictated by the equivalence principle,  the non-trivial geometry has physical consequences as can clearly be seen from the appearance of tidal forces in the geodesic deviation equation.

From the above discussion it is unequivocal that the Levi-Civita connection $\Gamma^\mu{}_{\alpha\beta}$ encodes both inertial and gravitational aspects of motion. From the viewpoint of local field theory, this dual role obscures the identification of gauge-invariant gravitational field degrees of freedom: Christoffel symbols depend on coordinates while physical observables should be tensorial. Earlier gauge approaches separated connection degrees of freedom and allowed torsion and spin to appear naturally in the gravitational sector; these approaches established the viability of gauge formulations as a route to reconciling gravity with the language of particle physics. \cite{Hehl1976,Blagojevic2002,Andringa2011}.

\begin{table}[h]
\centering
\renewcommand{\arraystretch}{1.4}
\begin{tabular}{@{}lcc@{}}
\toprule
\textbf{} & \textbf{Inertial Forces} & \textbf{Gravity} \\
\midrule

\textbf{In some coordinates} 
 & $g_{\mu\nu}=\eta_{\mu\nu}$ 
 & $g_{\mu\nu|p}= \eta_{\mu\nu}$ \\

\textbf{Force on test particles} 
 & $\displaystyle \mathrm{F}^\mu_{\!i} = -\widehat{\Gamma}^\mu{}_{\alpha\beta}\,
    \frac{dx^\alpha}{d\tau}\frac{dx^\beta}{d\tau}$
 & $\displaystyle \mathrm{F}^\mu_{\!g} = -\Gamma^\mu{}_{\alpha\beta}\,
    \frac{dx^\alpha}{d\tau}\frac{dx^\beta}{d\tau}$ \\

\textbf{Geodesic deviation} 
 & $\displaystyle \frac{\mathrm{D}^2\xi^\mu}{d\tau^2} = 0$
 & $\displaystyle \frac{\mathrm{D}^2\xi^\mu}{d\tau^2}
      = R^\mu{}_{\nu\alpha\beta}\,
        \frac{dx^\nu}{d\tau}\frac{dx^\alpha}{d\tau}\,\xi^\beta$ \\
\bottomrule
\end{tabular}
\caption{Comparison between inertial effects and gravitational effects in general relativity.}
\end{table}

In this work we follow this programme but take a different, constructive tack: we show that there is a reference (inertial) connection $\widehat{\Gamma}^\mu{}_{\alpha\beta}$ and the physical content of gravity is contained in the tensorial difference
\begin{equation}
\mathrm{K}^\mu{}_{\alpha\beta} \;=\; \Gamma^\mu{}_{\alpha\beta} - \widehat{\Gamma}^\mu{}_{\alpha\beta},
\end{equation}
which is coordinate-independent and therefore a natural candidate for the local gravitational field. From the equations of motion of test particles, it follows that \(\widehat\Gamma_{\alpha}\) is curvature free. To uniquely determine it, we extend infinitesimal translations to curved spacetime via vector fields $\xi_a{}^\mu$. The 1-forms \(\xi^{a}{}_{\mu}\) dual to these vector fields serve as gravitational potentials; the corresponding gauge-invariant field strength $\mathfrak{F}^{\xi}{}_{\,\mu\nu}^a$ is a Yang–Mills curvature for \(\mathrm{SU(2)\times U(1)}\) gauge theory. The idea that gravity may admit an interpretation in terms of gauge potentials and field strengths is both conceptually attractive and supported by multiple formalisms in the literature. \cite{Mansouri1976,Trautman1973,Blagojevic2002}.

A scalar field $\phi^{2}$ is introduced to capture the Newtonian potential and to ensure the correct weak–field limit. The combined action we construct contains Yang–Mills–like kinetic terms for $\xi^a{}_\mu$, a minimal coupling to $\phi$, and potential terms that control deviations from GR. In the limit of vanishing gravitational coupling parameter $\mathfrak{g}\to0$ the theory reduces to GR (the Einstein–Hilbert sector is recovered), while for finite $\mathfrak{g}$ the extra degrees of freedom lead to phenomenology that can naturally address cosmological puzzles such as accelerated expansion and modified galactic dynamics. This strategy is in line with other modern attempts to relate gauge dynamics and gravity (including amplitude-level relations such as the double-copy), while remaining fully classical and local in its construction. \cite{Bern2010,BCJReview2019,Clifton2012}.

The remainder of the paper is organized as follows. Section~2 distinguishes gravitational and inertial forces and introduces the tensor $\mathrm{K}^\mu{}_{\alpha\beta}$. Section~3 revisits the notion of momentum and its scalar character. Section~4 extends translations to curved spacetime through the vector fields $\xi_a{}^\mu$. Section~5 formulates the gravitational field strength tensor, while Section~6 demonstrates that gravity is an $\mathrm{SU(2)\times U(1)}$ gauge theory. Sections~7 and~8 establish the Newtonian limit and derive the corresponding field equations. Section~9 shows that conventional gauge interactions arise as higher-dimensional limits of the theory, and Section~10 discusses phenomenological implications for dark energy and dark matter. Finally, we summarize and outline directions for future work in Section~11.

\section{Gravity: A Force and Curved Spacetime}
To substantiate our assertion that gravity constitutes a genuine force, we begin by examining the case of electromagnetism, which is universally acknowledged as a fundamental interaction. By applying to electromagnetism the same line of reasoning that, in general relativity (GR), leads to the conclusion that particles in a gravitational field follow geodesics and are therefore “free,” we are seemingly driven to the analogous conclusion that electromagnetism is not a force. This apparent paradox reveals that such a conclusion arises solely from an inappropriate definition of acceleration. Once the notion of acceleration is correctly formulated in the context of GR, it follows that gravity must be regarded as a genuine force on the same footing as other fundamental interactions.

{Except in Section~9, we adopt the following convention:}
\begin{itemize}
  \item Latin indices \( a, b, c, \dots, h \) are label indices ranging from \( 0 \) to \( 3 \).
  \item Indices \( i, j, k, \dots \) range from \( 1 \) to \( 3 \) and denote coordinate indices, except when enclosed in brackets, in which case they serve as labels.
  \item Greek indices \( \mu, \nu, \dots \) are always used as coordinate indices.
\end{itemize}

In arbitrary coordinates, the equation of motion for a point particle in an electromagnetic field is
\begin{equation}\label{LorentzFOrce}
    \frac{d^{2}x^{\mu}}{d\tau^{2}} + \widehat\Gamma^{\mu}{}_{\lambda\nu}\frac{dx^{\lambda}}{d\tau}\frac{dx^{\nu}}{d\tau} =\frac{e}{m}\mathrm{F}^{\mu}{}_{\nu}\frac{dx^{\nu}}{d\tau},
\end{equation}
where the term \(\widehat\Gamma^{\mu}{}_{\lambda\nu}\frac{dx^{\lambda}}{d\tau}\frac{dx^{\nu}}{d\tau} \) represents a fictitious force. We show that by starting from non-covariant definition of acceleration, we are let to the conclusion that electromagnetism is a fictitious force. If we naively assume particles following geodesics of Levi-Civita connection are free we end up with a covariant definition of acceleration that suggests that particles in the electromagnetic field are free. Let us introduce a non-covariant acceleration
\[
\widehat a^{\mu} =\frac{d^{2}x^{\mu}}{d\tau^{2}}
\]
and assume that the charge-to-mass ratio satisfies
\[
\frac{e}{m} =\mathrm{A}_{\mu}\frac{dx^{\mu}}{d\tau}.
\]
We can always choose \(\mathrm{A}_{\mu}\) such that this is true at some point by exploiting the fact that it is gauge dependent. 
If \(\widehat a^{\mu} \) is naively identified with the acceleration in Newton’s second law, then the electromagnetic field strength can be written as
\begin{equation}
    \widehat{\mathrm{F}}_{\rho}/m =- \left(\widehat{\Gamma}_{\rho\lambda\nu} -\frac{1}{2}\left(\mathrm{A}_{\lambda}\mathrm{F}_{\rho\nu}+\mathrm{A}_{\nu}\mathrm{F}_{\rho\lambda}+\mathrm{A}_{\rho}\mathrm{F}_{\lambda\nu} \right) \right)\frac{dx^{\lambda}}{d\tau}\frac{dx^{\nu}}{d\tau}.
\end{equation}
In flat spacetime, the connection \(\widehat\Gamma^{\mu}{}_{\lambda\nu} \) takes the form
\begin{equation}\label{ficConnection}
    \begin{aligned}
    \widehat\Gamma_{\rho\lambda\nu} &= \partial_{\rho}x^{a}\,\partial_{\lambda}\partial_{\nu}x_{a}\\
    &=\frac{1}{2}\left(\partial_{\lambda}\widehat g_{\rho\nu}+\partial_{\nu}\widehat g_{\rho\lambda}-\partial_{\rho}\widehat g_{\lambda\nu} \right),
    \end{aligned}
\end{equation}
where \(\widehat g_{\mu\nu}= \partial_{\mu}x^{a}\partial_{\nu}x_{a} \) is the flat-spacetime metric expressed in arbitrary coordinates. On this basis, we introduce the Christoffel symbols \(\Gamma^{\mu}{}_{\lambda\nu} \) associated with the effective metric
\begin{equation}\label{electromagnetic spacetime}
    g_{\mu\nu} =\widehat g_{\mu\nu} + \mathrm{A}_{\mu}\mathrm{A}_{\nu},
\end{equation}
and we can rewrite the electromagnetic field strength vector as
\begin{equation}
    \widehat{\mathrm{F}}^{\mu}/m = -\Gamma^{\mu}{}_{\lambda\nu}\frac{dx^{\lambda}}{d\tau}\frac{dx^{\nu}}{d\tau}.
\end{equation}
Since there always exists a local frame in which \(\Gamma^{\mu}{}_{\lambda\nu} \) can be made to vanish at a point, it would then follow that \(\widehat{\mathrm{F}}^{\mu} \) is a fictitious force. However, unlike purely inertial fictitious forces (whose connection coefficients are given by \eqref{ficConnection}), the quantity \(\widehat{\mathrm{F}}^{\mu} \) encapsulates certain genuine electromagnetic effects—specifically, the relative deviation of neighboring trajectories of charged particles—through the non-vanishing curvature associated with \(\Gamma^{\mu}{}_{\lambda\nu} \).

Thus, the inference that the electromagnetic force is fictitious is seen to originate from the misuse of a non-covariant definition of acceleration. To recover the physically correct interpretation, in which electromagnetism is recognized as a genuine interaction, one must employ a covariant definition of acceleration. If we adopt the GR definition:
\begin{equation}
    a_{\mathrm{GR}}^{\mu} = \frac{d^{2}x^{\mu}}{d\tau^{2}} +\Gamma^{\mu}{}_{\lambda\nu}\frac{dx^{\lambda}}{d\tau}\frac{dx^{\nu}}{d\tau},
\end{equation}
then for the motion of particles in the electromagnetic field as encoded in the metric \eqref{electromagnetic spacetime}, we obtain \(a_{\mathrm{GR}}^{\mu} =0\). Within this formalism, one would be led to conclude that electromagnetism is not a force but rather a manifestation of spacetime curvature. To restore the standard view that electromagnetism is a genuine force, it is therefore essential to employ the correct definition of acceleration given in \eqref{Acceleration definition}.

Since bona fide Cartesian coordinates cannot be introduced globally in a curved spacetime, and quasi-Cartesian coordinates are in general non-unique, the condition \(g^{\mu\nu}\neq \widehat g^{\mu\nu}\) (which is manifestly satisfied in a \(4\mathcal{D}\) spacetime) implies that the connection coefficients \(\widehat\Gamma_{\rho\lambda\nu}\) are not uniquely determined (a procedure to construct them uniquely will be presented in the following sections). Moreover, the inequality \(g^{\mu\nu}\neq \widehat g^{\mu\nu}\) entails that equation \eqref{LorentzFOrce} represents the Lorentz force law in a genuinely curved spacetime.  

If, instead, we extend the dimensionality of spacetime to five and impose the condition \(\partial_{5}x^{a}=0\), then the \(4\mathcal{D}\) spacetime sector of the contravariant metric coincides with the flat Minkowski metric. In this higher-dimensional setting, the coefficients \(\widehat\Gamma_{\rho\lambda\nu}\) become uniquely defined. One then identifies the \(5\mathcal{D}\) extension of \(g_{\mu\nu}\) precisely with the Kaluza–Klein metric. Consequently, Kaluza–Klein theories can be viewed as emerging from an initially inappropriate definition of acceleration. Table~1 shows that this is also the case with GR.

The assertion that particles following curves of extremal arc length (i.e., geodesics of the Levi-Civita connection) are not necessarily free may appear counterintuitive at first. Nonetheless, once one recalls that the extremality of the arc length depends on the chosen metric, it becomes clear that, by a suitable choice of metric, essentially any given curve can be rendered an extremal one.

For example, with respect to the metric \(\widehat g_{\mu\nu}\), the curve determined by the variational principle
\[
    \mathrm{S} = \int \sqrt{\widehat g_{\mu\nu} \, dx^{\mu} dx^{\nu}} \quad \text{with} \quad \delta \mathrm{S} = 0
\]
has extremal arc length. By contrast, the curves determined by
\[
    \mathrm{S_{1}} = \int \sqrt{\phi^{2}(x)\,\widehat g_{\mu\nu} \, dx^{\mu} dx^{\nu}} 
    \quad \text{with} \quad \delta \mathrm{S_{1}} = 0
\]
and
\[
    \mathrm{S_{2}} = \int \sqrt{\widehat g_{\mu\nu} \, dx^{\mu} dx^{\nu} 
    + \frac{e}{m}\,\mathrm{A}_{\mu} dx^{\mu} d\tau} 
    \quad \text{with} \quad \delta \mathrm{S_{2}} = 0, \quad 
    \text{and} \quad \frac{e}{m} d\tau = \mathrm{A}_{\mu} dx^{\mu}
\]
are not, in general, extremal with respect to \(\widehat g_{\mu\nu}\). However, these same curves are geodesics (i.e., curves of extremal arc length) with respect to the modified metrics
\[
    g^{(1)}_{\!\mu\nu} = \phi^{2}(x)\,\widehat g_{\mu\nu}, 
    \qquad
    g^{(2)}_{\!\mu\nu} = \widehat g_{\mu\nu} + \mathrm{A}_{\mu}\mathrm{A}_{\nu},
\]
respectively. This illustrates that the designation of a trajectory as “free” in the sense of being a geodesic crucially depends on the underlying metric structure of the spacetime.

The definition of acceleration that accounts for inertial effects is
\begin{equation}\label{Acceleration definition}
a^\mu = \frac{d^2x^\mu}{d\tau^2} + \widehat\Gamma^\mu{}_{\alpha\beta} \frac{dx^\alpha}{d\tau} \frac{dx^\beta}{d\tau}.
\end{equation}
Re-expressing equations of motion for test particles in a gravitational field
$$\frac{d^2x^\mu}{d\tau^2} + \Gamma^\mu{}_{\alpha\beta} \frac{dx^\alpha}{d\tau} \frac{dx^\beta}{d\tau} = 0$$
in terms of this, we get
$$a^\mu = - (\Gamma^\mu{}_{\alpha\beta} - \widehat\Gamma^\mu{}_{\alpha\beta}) \frac{dx^\alpha}{d\tau} \frac{dx^\beta}{d\tau}.$$
Thus, the gravitational field strength is
\begin{equation}\label{GravitationalFieldStrength}
\mathrm{F}_{g}{}^\mu/m = -\mathrm{K}^\mu{}_{\alpha\beta} \frac{dx^\alpha}{d\tau} \frac{dx^\beta}{d\tau}
\end{equation}

where $\mathrm{K}^{\mu}{}_{\alpha\beta} = \Gamma^\mu{}_{\alpha\beta} - \widehat\Gamma^\mu{}_{\alpha\beta}$.

$\mathrm{K}^\mu{}_{\alpha\beta}$ is the difference between two connections, thus it is a tensor and cannot be transformed away by a change of coordinates. This shows that gravity is a real force.

Let $\Gamma_{\alpha}$ be the connection 1-form with connection coefficients $\Gamma^\mu_{\alpha\beta}$. The curvature two-form is

$$R_{\alpha\beta} = [\partial_{\alpha} +\Gamma_\alpha,\partial_{\beta} + \Gamma_\beta].$$
If $\Gamma_{\alpha}$ is associated with the Levi-Civita connection, the coefficients $R^\mu{}_{\nu\alpha\beta}$ of the curvature 2-form are components of the Riemann curvature tensor. Since the physical content of gravity is encapsulated in the Riemann curvature tensor, if we take the candidate for the gravitational field strength tensor to be $\mathrm{K}^{\mu}{}_{\alpha\beta}$, it must be possible to write the Riemann curvature tensor in terms of $\mathrm{K}^{\mu}{}_{\alpha\beta}$. A straightforward calculation shows that if we introduce the 1-form $\mathrm{K}_{\alpha}$ whose coefficients are $\mathrm{K}^{\mu}{}_{\alpha\beta}$, we can write the curvature 2-form as
\begin{equation}
 R_{\alpha\beta} = [\widehat\nabla_{\alpha},\mathrm{K}_{\beta} ]- [\widehat\nabla_{\beta},\mathrm{K}_{\alpha}] + [\mathrm{K}_{\alpha}, \mathrm{K}_{\beta}]
\end{equation}
thus we can write the Riemann curvature tensor entirely in terms of $\mathrm{K}^{\mu}{}_{\alpha\beta}$.

It also follows from table 1 that the inertial effects are characterized by the 1-form with vanishing curvature. Therefore, the general form of $\widehat\Gamma_{\alpha}$ is
\begin{equation}
 \widehat\Gamma_{\alpha} = \xi^{-1}(x)\partial_{\alpha}\xi(x)
\end{equation}
where $\xi$ is an invertible $4\times 4$ matrix.
We can use this to rewrite acceleration as
\begin{equation}\label{Acceleration From Momentum}
 m a^{\mu} = \xi_{a}{}^{\mu}\frac{d\mathrm{P}^{a}}{d\tau}
\end{equation}
where the scalar $\mathrm{P}^{a}= m \xi^{a}{}_{\mu}\frac{dx^{\mu}}{d\tau}$ is momentum, $\xi^{a}{}_{\mu}$ are components of the matrix $\xi$, and $\xi_{a}{}^{\mu}$ the components of $\xi^{-1}$. It should be noted that we can not just take \(\xi^{a}{}_{\mu}=\partial_{\mu}x^{a} \) since the Cartesian coordinates do not exist in curved spacetime and the quasi-cartesian coordinates are not unique.

\section{Revisiting Momentum}
That \(\{\mathrm{P}^{a} \} \) are momenta is clear from the fact that they are conserved in the absence of a force. Since the fact that they are scalars might come as a surprise given we usually think of momentum as a vector, it is worth showing that this makes sense.

Recall that the energy–momentum tensor in field theory provides the conserved Noether current associated with spacetime translations~\cite{Wald1984,Weinberg1995}. 
That is momenta arise as charges corresponding to infinitesimal translations generated by the vector fields $\xi_a{}^\mu$. When these infinitesimal translations are the symmetries of spacetime, it is clear  that \(\{\mathrm{P}_{a}\}\)  are conserved for particles following geodesics of Levi-Civita connection.Thus it makes sense to think of them as the scalar Noether charges linked to the translational symmetry group~\cite{Trautman1973,Blagojevic2002}. We show that this is indeed the case starting from the definition of momentum in terms of the energy-momentum tensor.

Given the energy–momentum tensor \(\mathrm{T}^{\mu\nu}\) in flat spacetime expressed in Cartesian coordinates, the momentum is defined as  
\begin{equation}\label{Cartesian-Momentum}
    \mathrm{P}^{\mu} \equiv \int \mathrm{T}^{\mu}{}_{\nu}\,\star dx^{\nu},
\end{equation}
and its conservation follows from \(\partial_{\mu}\mathrm{T}^{\mu\nu} = 0\). However, the definition \eqref{Cartesian-Momentum} does not yield a genuine vector in general, since the right-hand side does not transform as a vector under arbitrary coordinate transformations. Consequently, \eqref{Cartesian-Momentum} cannot be regarded as a generally covariant definition of momentum.

To obtain an appropriate, coordinate-independent definition of momentum from the energy–momentum tensor, the integrand in \eqref{Cartesian-Momentum} must be invariant under general coordinate transformations. This requirement can be satisfied by defining momentum in arbitrary coordinates as  
\begin{equation}\label{vector-Momentum}
    \mathrm{P}^{\mu} = \xi_{a}{}^{\mu} \int \mathfrak{J}^{a}{}_{\nu}\,\star dx^{\nu},
\end{equation}
where \(\mathfrak{J}^{a}{}_{\mu} \equiv \xi^{a\nu}\mathrm{T}_{\nu\mu}\), and in flat spacetime with Cartesian coordinates one has \(\xi_{b}{}^{\mu} = \Lambda_{b}{}^{a}\delta_{a}{}^{\mu}\) for some global Lorentz transformation \(\Lambda\).

For a point source, we identify \(\mathrm{P}^{\mu} = m\,dx^{\mu}/d\tau\). Using \eqref{Acceleration definition}, we infer that momentum is conserved (\(a^{\mu} = 0\)) when \(\partial_{\mu}\mathfrak{J}^{a\mu} = 0\), and this conservation law is manifestly independent of the choice of coordinates. From \eqref{vector-Momentum} it follows that  
\[
\mathrm{P}^{a} = \xi^{a}{}_{\mu}\,\mathrm{P}^{\mu}
\]
is a Noether charge associated with the infinitesimal transformation \(\delta x^{\mu} = \epsilon\,\xi_{a}{}^{\mu}\), which, in Cartesian coordinates (given the form of \(\xi_{a}{}^{\mu}\)), is clearly identified as a translation. Therefore, the components \(\mathrm{P}^{a}\), being Noether charges, are scalar quantities with respect to general coordinate transformations (while transforming as a Lorentz vector under changes of the internal index \(a\)).

To justify the conclusion that the momenta defined by \eqref{Acceleration From Momentum} are scalars, starting from the energy–momentum tensor, we assumed that in Cartesian coordinates
\[
\xi_{b}{}^{\mu} = \Lambda_{b}{}^{a}\,\delta_{a}{}^{\mu}.
\]
We now demonstrate that this assumption indeed holds.

In flat spacetime, we have \(\mathrm{K}^{\mu}{}_{\alpha\beta} = 0\). Hence \(\widehat{\Gamma}^{\mu}{}_{\alpha\beta}\) coincides with the Levi-Civita connection (Christoffel symbols) associated with the flat metric. From the metric compatibility condition \(\widehat{\nabla}_{\mu} g_{\alpha\beta} = 0\), where \(g_{\alpha\beta}\) is the flat spacetime metric written in arbitrary coordinates, we obtain
\begin{equation}\label{MetricInxi}
    g_{\mu\nu} = \eta_{ab}\,\xi^{a}{}_{\mu}\,\xi^{b}{}_{\nu}.
\end{equation}
Imposing the torsion-free condition implies that, at least locally, the fields \(\xi^{a}{}_{\mu}\) can be written as pure gradients:
\begin{equation}\label{flat xi}
    \xi^{a}{}_{\mu} = \partial_{\mu}\psi^{a}(x).
\end{equation}
If \(\{x^{\mu}\}\) are Cartesian coordinates, then the flatness of the metric and the form \eqref{MetricInxi} require
\[
\psi^{a}(x) = \Lambda^{a}{}_{b}\,x^{b} + \alpha^{a},
\]
for some constant vector \(\alpha^{a}\) and constant Lorentz matrix \(\Lambda^{a}{}_{b}\). This establishes the validity of our assumption \(\xi_{b}{}^{\mu} = \Lambda_{b}{}^{a}\delta_{a}{}^{\mu}\) in Cartesian coordinates.

\section{Extension of Translations to Curved Spacetime}

Since \(\xi_{a}{}^{\mu}\) are the generators of spacetime translations, they are uniquely determined up to a global Lorentz transformation acting on the internal index \(a\). In contrast to tetrads, this residual freedom is purely global and thus the vector fields \(\xi_{a}{}^{\mu}\) possess a direct physical interpretation. The fact that the scalar functions \(\psi^{a}\) are uniquely defined only up to a global Poincaré transformation is therefore fully consistent with the characterization of the one-forms dual to the generators of translations in flat spacetime provided by \eqref{MetricInxi} and \eqref{flat xi}.

When \(\mathrm{K}^{\mu}{}_{\alpha\beta} \neq 0\), the connection coefficients associated with fictitious (inertial) forces become distinct from the Christoffel symbols of the metric. Moreover, one can no longer introduce global Cartesian coordinates; instead, one may construct quasi-Cartesian (Riemann normal) coordinates in a neighborhood of a point. These coordinates are not unique. Specifically, if \(\{x^{\mu}\}\) are Riemann normal coordinates about some point, then another set of Riemann normal coordinates \(\{y^{\mu}\}\) at the same point can be related to \(\{x^{\mu}\}\) by
\begin{equation}
    y^{\mu} = a^{\mu} + \Lambda^{\mu}{}_{\nu} x^{\nu}
    + \frac{1}{6}\,\mathrm{C}^{\mu}{}_{\alpha\beta\gamma}\,x^{\alpha}x^{\beta}x^{\gamma}
    + \mathcal{O}(x^{4}),
\end{equation}
where \(a^{\mu}\) is a constant translation vector, \(\Lambda^{\mu}{}_{\nu}\) is a Lorentz transformation, and \(\mathrm{C}^{\mu}{}_{\alpha\beta\gamma}\) encodes higher-order curvature-dependent corrections. This transformation law makes explicit the non-uniqueness of Riemann normal coordinates while preserving the local “quasi-Cartesian’’ structure at the chosen point. In particular, if \(\{\psi^{a}\}\) and \(\{\tilde{\psi}^{a}\}\) are each defined in terms of Riemann normal coordinates, then they are related by
\[
\tilde{\psi}^{a} = \Lambda^{a}{}_{b}\,\psi^{b} + \alpha^{a}(x),
\]
where
\[
\alpha^{a}(x) = \tfrac{1}{6}\,\mathrm{C}^{a}{}_{\alpha\beta\gamma}\,x^{\alpha}x^{\beta}x^{\gamma} + \mathcal{O}(x^{4}).
\]
Consequently, in this setting the definition of \(\xi^{a}{}_{\mu}\) given by \eqref{flat xi} is no longer compatible with the interpretation of \(\xi_{a}{}^{\mu}\) as generators of infinitesimal translations. The reason is that now any two fields \(\xi^{a}{}_{\mu}\) related by
\[
\xi^{a}{}_{\mu} \longmapsto \xi^{a}{}_{\mu} + \partial_{\mu}\alpha^{a}
\]
are physically equivalent, which contradicts the requirement that \(\xi_{a}{}^{\mu}\) uniquely generate translations.

This issue can be resolved by introducing a vector field \(\mathrm{A}^{a}{}_{\mu}\) defined such that, whenever
\[
\tilde{\psi}^{a} = \Lambda^{a}{}_{b}\psi^{b} + \alpha^{a},
\]
it transforms according to
\[
\tilde{\mathrm{A}}^{a}{}_{\mu} = \Lambda^{a}{}_{b}\,\mathrm{A}^{b}{}_{\mu} + \partial_{\mu}\alpha^{a}.
\]
Defining the 1-forms dual to the translation generators by
\begin{equation}\label{translation 1-forms}
    \xi^{a}{}_{\mu} = \partial_{\mu}\psi^{a} - \mathrm{A}^{a}{}_{\mu}
\end{equation}
then restores consistency with the requirement that the set \(\{\xi_{a}{}^{\mu}\}\) indeed represents generators of translations.

An alternative route to \eqref{translation 1-forms} is to impose that the connection \(\widehat{\nabla}_{\mu}\) be metric compatible. Demanding that \(\widehat{\nabla}_{\mu}\) differ from the Levi–Civita connection then automatically implies the presence of torsion, which in turn leads to the relation \eqref{translation 1-forms}. One would subsequently need to justify why \(\widehat{\nabla}_{\mu}\) must be metric compatible. Within this framework, metric compatibility emerges naturally from the requirement that \(\{\xi_{a}{}^{\mu}\}\) represent the extension of infinitesimal translations to curved spacetime.

Equation \eqref{translation 1-forms} also makes it evident that \(\xi^{a}{}_{\mu}\) closely resembles the 1-form appearing in Poincaré gauge theories of gravity~\cite{Kibble1961,Hehl1976}, which serves as the gauge potential associated with the translational sector of the Poincaré group~\cite{Blagojevic2002,BlagojevicHehl2013}. However, \(\xi^{a}{}_{\mu}\) is not itself a translation gauge potential, since extending translations to curved spacetimes does not amount to gauging the Poincaré group. This is manifest in the fact that the vector fields \(\xi_{a}{}^{\mu}\) generating infinitesimal translations in curved spacetime do not commute and therefore do not furnish a representation of the translational part of the Lie algebra of the Poincaré group.

\section{The Gravitational Field Strength Tensor}   
We can rewrite \(\mathrm{K}^{\mu}{}_{\alpha\beta} \) in terms of \(\xi^{a}{}_{\mu} \) as
\begin{equation}
   \mathrm{K}^{\mu}{}_{\alpha\beta} = 1/2\xi_{a}{}^{\mu} \left(\mathrm{F}^{a}{}_{\alpha\beta} +\mathscr{L}_{\xi^{a}}g_{\alpha\beta} \right) 
\end{equation}
where \[\mathrm{F}^{a}{}_{\mu\nu} = \partial_{\mu}\xi^{a}{}_{\nu}-\partial_{\nu}\xi^{a}{}_{\mu} \]
and \(\mathscr{L}_{\xi^{a}} \) is the Lie derivative of the metric along the integral curves of the vector field \(\xi^{a\mu} \). Clearly, we can write the gravitational field strength tensor as 
\begin{equation}\label{RealGravitFieldStrength}
    \mathrm{F}_{g}{}^{\mu}/m = 1/2\xi_{a}{}^{\mu}\mathscr{L}_{\xi^{a}}g_{\alpha\beta}\frac{dx^{\alpha}}{d\tau}\frac{dx^{\beta}}{d\tau}.
\end{equation}
From this it follows that in the absence of gravity we have
\begin{equation}\label{absentGravity}
    \mathscr{L}_{\xi^{a}}g_{\alpha\beta} = 0.
\end{equation}
Notice that in the absence of gravity \(\{\xi^{a}{}_{\mu}\} \) are not necessarily trivial since
in general
\begin{equation}
    \mathrm{F}^{a}{}_{\mu\nu}\neq 0.
\end{equation}
From \eqref{absentGravity} it is clear that in the absence of gravity \(\{\xi_{a}{}^{\mu} \} \) are Killing vector fields; as such, they form a closed Lie algebra. The Lie bracket in this case is
\begin{equation}\label{xiLieBracket}
    \xi_{a}{}^{\mu}\partial_{\mu}\xi_{b}{}^{\lambda}-\xi_{b}{}^{\mu}\partial_{\mu}\xi_{a}{}^{\lambda} =\mathfrak{gC}_{ab}{}^{c}\xi_{c}{}^{\lambda}
\end{equation}
where $\{\mathfrak{gC}_{ab}{}^{c}\} $ are structure constants and \(\mathfrak{g}\) is the new gravitational coupling constant distinct from Cavendish constant. By making use of the fact that the matrix $\xi$ is invertible, we can write \eqref{xiLieBracket} as
\begin{equation}
    \mathrm{F}^{a}{}_{\mu\nu}= -\mathfrak{gC}_{bc}{}^{a}\xi^{b}{}_{\mu}\xi^{c}{}_{\nu}.
\end{equation}
Let
\begin{equation}
    \mathfrak{F}^{\xi a}{}_{\mu\nu} = \partial_{\mu}\xi^{a}{}_{\nu}-\partial_{\nu}\xi^{a}{}_{\mu}+\mathfrak{gC}_{bc}^{a}\xi^{b}{}_{\mu}\xi^{c}{}_{\nu}.
\end{equation}
The absence of gravity is characterized by 
\begin{equation}
    \mathfrak{F}^{\xi a}{}_{\mu\nu} =0.
\end{equation}
We can rewrite \eqref{RealGravitFieldStrength} in terms of \(\mathfrak{F}^{\xi a}{}_{\mu\nu} \) as
\begin{equation}
    \mathrm{F}_{\! g}{}^{\mu} = \mathrm{P}_{b}\mathfrak{F}^{\xi b\mu}{}_{\nu}\frac{dx^{\nu}}{d\tau}.
\end{equation}
Thus \(\mathfrak{F}^{\xi a\mu}{}_{\nu} \) is the gravitational field strength tensor and we recognize \(\{\mathrm{P}_{a} \} \) as the gravitational charges. The gravitational field strength in our case is clearly different from the one proposed in either the Poincaré gauge theories of gravity, teleparallel gravity~\cite{Bahamonde:2021gfp} or in \(\mathrm{SL(2,\mathbb{C})}\) gauge theory of gravity\cite{Mansouri1976}.

\section{Gravity as the $\mathrm{SU(2)\times U(1)}$ Gauge Theory}
The gravitational field strength tensor \(\mathfrak{F}^{\xi a}{}_{\mu\nu} \) is precisely the field strength tensor of Yang-Mills theories with \(\xi^{a}{}_{\mu} \) as the potentials, implying that gravity can be viewed as a gauge theory. We assert that \(\{\xi^{a}{}_{\mu}\} \) are physical quantities, which means they must be gauge invariant. This seems to contradict the interpretation of gravity as a gauge theory, but we demonstrate that this is not the case.

Our assertion is that \(\{\xi^{a}{}_{\mu}\} \) are gauge invariant, which also makes \(\mathfrak{F}^{\xi a}{}_{\mu\nu} \) gauge invariant. We prove this by construction. We introduce the field \(\mathrm{V} \) and the gauge potential \(\mathcal{A}^{}_{\mu}\) with the gauge transformation \(\mathrm{U}\) behaving as:
\[\mathrm{V} \longmapsto \mathrm{VU} \]
and
\[\mathcal{A}{}_{\mu}\longmapsto \mathrm{U}^{-1}(\mathcal{A}_{\mu} + \partial_{\mu})\mathrm{U} \]
respectively.

By defining \(\xi_{\mu} = \tau_{a}\xi^{a}{}_{\mu} \) (where \(\{\tau_{a}\}\) form the matrix representation of the Lie algebra of the gauge group considered) as:
\[
    \xi_{\mu} = \mathrm{V}(\mathcal{A}_{\mu} + \partial_{\mu})\mathrm{V}^{-1}
\]
we observe that it is indeed gauge invariant. This construction of the gauge invariant potential mirrors the Stueckelberg procedure~\cite{Stueckelberg:1938hvi,kunimasa1967generalization}.

Having confirmed that interpreting gravity as a gauge theory fits within this framework, we identify the gauge group. From the previous section, it is clear that the manifold corresponding to the vacuum solution is the group manifold associated with the relevant group. Since there are no 4-dimensional (semi)simple groups, and the groups associated with the vacuum solution comprise the non-Abelian groups, the group manifold is the product manifold \(\mathcal{M}_{3}\times\mathbb{R} \). Since in vacuum we expect no preferred spatial direction, \(\mathcal{M}_{3}=\mathrm{SU(2)} \). Due to the fact that the formalism gives us information regarding only the local properties of the manifold, we may take the group manifold to be \(\mathrm{SU(2)\times U(1)} \). The other group consistent with the vacuum solutions is the group of translations. This is readily obtained from \(\mathrm{SU(2)} \) by group contraction which is essentially setting the gravitational coupling constant \(\mathfrak{g}\) to zero. 

\section{The Newtonian Potential}
In general relativity, the non-trivial solutions to the vacuum equations -under the assumption that spacetime is spherically symmetric and static is the Schwarzschild solution which reproduces Newtonian gravity for a point source in the non-relativistic limit. There exists the coordinates (Schwarzchild coordinates) in which the metric has the form
\begin{subequations}
    \begin{align}
    g_{00}=-(1-2\mathscr{G}M/r) & & g_{ij} = \delta_{ij}+\frac{2\mathscr{G}M/r^{3}}{1-2\mathscr{G}M/r}x_{i}x_{j}.
    \end{align}
\end{subequations}
We show that in the current state of the formalism where gravity is completely described by \(\{\xi^{a}{}_{\mu}\} \) such a solution does not exist.

Let us assume that gravity is completely described by \(\{\xi^{a}{}_{\mu}\} \). There existence of time translation symmetry implies that 
\begin{equation}
    \frac{d\mathrm{P}^{(0)}}{d\tau} = 0.
\end{equation}
We can always choose a coordinate system in which \[\xi_{(0)}{}^{\mu} =1/\mathrm{B}(r) \delta_{0}{}^{\mu} \]
and all \(\{\xi^{a}{}_{\mu}\}\) are time independent.
Let \(\xi^{(0)}{}_{\mu} =\mathrm{B}_{\mu} \) with \(\mathrm{B}_{0} = \mathrm{B}(r). \) In such a coordinate system
\begin{equation}
    \xi^{a}{}_{\mu} = \delta^{a}{}_{i}\xi^{(i)}{}_{j}\delta^{j}_{\mu}+\delta^{a}{}_{0}\mathrm{B}_{\mu}.
\end{equation}
As such, the metric is 
\begin{equation}
    g_{\mu\nu} =\delta^{i}_{\mu}\delta^{j}{}_{\nu}\widehat{g}_{ij}-\mathrm{B}_{\mu}\mathrm{B}_{\nu}
\end{equation}
where \(\widehat{g}_{ij} =\xi^{(k)}{}_{i}\xi_{(k)j} \). Since the spacetime is static, we can further choose the coordinates such that \(\mathrm{B}_{i} =0\). Clearly, we can recover Newtonian gravity or its modification if \(\mathrm{B}(r)^{2} = 1-2\Phi \) where \(\Phi(r) \) reduces to the Newtonian potential in a certain limit.

Making use of the equation of motion of test particles in the gravitational field together with the above stated choices, we find that
\begin{equation}
    \frac{d\mathrm{P}_{(0)}}{d\tau} = -\mathrm{P}_{(0)}\frac{x_{i}}{r}\frac{dx^{i}}{d\tau}\partial_{r}\ln{(\mathrm{B}(r))}
\end{equation}
conservation of momentum in the temporal direction then implies that \(\mathrm{B}(r)\) is constant. Thus the theory in its current form does not reproduce Newtonian gravity. 

In order to recover Newtonian gravity, we have to introduce the Newtonian potential as the genuine gravitational degree of freedom supplementing \(\xi_{\mu} \). For later convenience we introduce the scalar field \(\phi\) and define the Newtonian potential (\(\Phi \)) by 
\begin{equation}
    \phi^{2} = 1+2\Phi.
\end{equation}
The gravitational force then becomes
\begin{equation}\label{CompleteGravitationalForce}
    \mathrm{F}_{g}{}^{\mu} = \mathrm{P}_{b}\mathfrak{F}^{\xi b\mu}{}_{\nu}\frac{dx^{\nu}}{d\tau}-\frac{m}{2}\partial^{\mu}\phi^{2}.
\end{equation}

While in this case we get Newtonian gravity, it is clear that \(g_{\alpha\beta} \) will not reduce to the solution to Einstein field equations even in the limit where the theory reproduces Newtonian gravity. This is expected, since we know that the solution to Einstein field equations is the metric of spacetime in which the trajectories of particles under the influence of the force given by equation \eqref{CompleteGravitationalForce} are geodesics. Such a solution \(\tilde{g}_{\alpha\beta}\) is related to \(g_{\alpha\beta} \) by 
\begin{equation}\label{conformal metrics}
   \tilde g_{\alpha\beta} = \phi^{2}g_{\alpha\beta}
\end{equation}
when the dynamics of \(\xi^{a}{}_{\mu} \) are equivalent to those of \(g_{\mu\nu} \).

\section{The Gravitational Field Equations}
From \eqref{CompleteGravitationalForce} it becomes evident that the equations of motion of test particles in the gravitational field can be expressed as follows:
\begin{equation}\label{trajectories of TestP}
    \frac{d^{2}x^{\mu}}{ds^{2}} + \Big\lbrace \Gamma^{\mu}{}_{\lambda\nu} -\mathfrak{g}/2\mathfrak{C}^{a}{}_{bc}\xi^{b}{}_{\lambda}\xi^{c}{}_{\nu} +\frac{\partial_{\rho}\phi^{2}}{\phi^{2}}\left(\delta^{\rho}{}_{(\lambda}\delta_{\nu)}{}^{\mu} -1/2 g^{\rho\mu}g_{\nu\lambda} \right) \Big\rbrace\frac{dx^{\lambda}}{ds}\frac{dx^{\nu}}{ds} =0
\end{equation}
by utilizing \(g_{\mu\nu}dx^{\mu}dx^{\nu}=-\phi^{2}d\tau^{2} \) and the definition \(ds =\phi^{2}d\tau \). We denote the expression in the curly braces in \eqref{trajectories of TestP} by \(\tilde\Gamma^{\mu}{}_{\lambda\nu} \). In scenarios where \(\mathfrak{g}\to 0\), \(\tilde\Gamma^{\mu}{}_{\lambda\nu} \) are the Christoffel symbols derived from the metric \(\tilde{g}_{\alpha\beta} \) define by \eqref{conformal metrics}. Although the middle term in the expression of \(\tilde\Gamma^{\mu}{}_{\lambda\nu} \) becomes inconsequential in the equations of motion of test particles in the gravitational field, it can not be dispensed with since it breaks both conformal invariance and local Lorentz invariance, whose presence would render \(\xi^{}_{\mu} \) nonphysical.

To derive the Lagrangian that describes the dynamics of the gravitational field, a connection \(\tilde\nabla_{\lambda} \) is introduced, defined by: 
\begin{equation}
    \tilde\nabla_{\lambda} = \partial_{\lambda} +\tilde\Gamma_{\lambda}.
\end{equation}
The coordinate invariant Lagrangian, constructed from this connection and at most quadratic in the derivatives of \(\xi^{a}{}_{\mu} \), is
\begin{equation}\label{Gravitation Lagrangian}
    \mathrm{L} = \int\sqrt{g}d^{3}x\frac{dt}{d\tau}\phi^2 \xi_{a}{}^{\mu}\left[\tilde\nabla_{\nu},\tilde\nabla_{\mu} \right]\xi^{a\nu}
\end{equation}
where \(dt=dx^{0}\). When \(\mathfrak{g}=0\), \(\tilde\nabla_{\mu}\) is torsion-free. As such, we have:
\begin{equation}\label{torsion free Comm}
    \left[\tilde\nabla_{\mu},\tilde\nabla_{\nu} \right](\Lambda
    ^{b}{}_{a}(x)\xi^{a\lambda} ) = \Lambda^{b}{}_{a}(x)\left[\tilde\nabla_{\mu},\tilde\nabla_{\nu} \right]\xi^{a\lambda}
\end{equation}
for any scalar fields \(\Lambda^{a}{}_{b}(x) \) and vector fields \(\xi^{a\nu} \). The theory clearly maintains invariance under local Lorentz transformation of \(\xi^{a}{}_{\mu}  \). The local transformation 
\begin{subequations}
    \begin{align*}
        \xi^{a}{}_\mu &\longmapsto \vartheta\xi^{a}{}_{\mu} \implies \sqrt{g}\longmapsto \vartheta^{4}\sqrt{g},\\
        \phi^{2} &\longmapsto\phi^{2}\vartheta^{-2},\\
        \frac{d}{d\tau} &\longmapsto \frac{1}{\vartheta^{2}}\frac{d}{d\tau},\\
        \tilde\nabla &\longmapsto \tilde\nabla
    \end{align*}
\end{subequations}
which leaves \eqref{trajectories of TestP} unchanged is also a symmetry of the action in this context. This indicates that when \(\mathfrak{g}\to 0 \) the theory can not differentiate between metrics related by a conformal transformation. Consequently, even when the coordinates are specified the metric is not unique in this scenario. If this theory corresponds to GR when \(\mathfrak{g}\to 0\), this highlights the ambiguity inherent in the definition of force in general relativity. On one hand the metric of spacetime is \(g_{\mu\nu} \) and there is a force derived from the scalar potential $\phi^{2}$ while on the other hand the metric of spacetime is \(\tilde g_{\mu\nu} \) and there is no force and there is no way of distinguishing the two situations within the general relativity framework.  If we make a choice $\vartheta=\phi^{-1}$, we find that the Lagrangian reduces to the Einstein Hilbert Lagrangian affirming that the theory is indeed GR when \(\mathfrak{g}\to 0\). 
When \(\mathfrak{g}\neq 0\), \eqref{torsion free Comm} is no longer valid. We have instead
\begin{equation}\label{torsion Comm}
    \left[\tilde\nabla_{\mu},\tilde\nabla_{\nu} \right](\Lambda
    ^{b}{}_{a}(x)\xi^{a\lambda} ) =\left(\left[\tilde\nabla_{\mu},\tilde\nabla_{\nu} \right]\Lambda
    ^{b}{}_{a}(x)\right)\xi^{a\lambda}  + \Lambda^{b}{}_{a}(x)\left[\tilde\nabla_{\mu},\tilde\nabla_{\nu} \right]\xi^{a\lambda}.
\end{equation}
Thus the above stated symmetries are broken. 

Let 
\begin{equation}
    \tilde{\mathrm{R}}_{\nu\rho} = \partial_{\mu}\tilde\Gamma^{\mu}{}_{\nu\rho}-\partial_{\nu}\tilde\Gamma^{\mu}{}_{\mu\rho}+\tilde\Gamma^{\delta}{}_{\nu\rho}\tilde\Gamma^{\mu}{}_{\mu\delta}-\tilde\Gamma^{\delta}{}_{\mu\rho}\tilde\Gamma^{\mu}{}_{\nu\delta}.
\end{equation}
We can write the Lagrangian density as
\begin{equation}
    \mathcal{L} =\sqrt{g}\phi^{2}g^{\nu\rho}\Big\lbrace\tilde{\mathrm{R}}_{\nu\rho}+\xi_{a}{}^{\mu}\tilde{\nabla}_{\delta}\xi^{a}{}_{\rho}\left(\tilde{\Gamma}^{\delta}{}_{\mu\nu}-\tilde{\Gamma}^{\delta}{}_{\nu\mu} \right) \Big\rbrace.
\end{equation}
Making use of the definition of \(\tilde\Gamma^{\delta}{}_{\mu\nu} \) with \(\mathrm{R}\) as the Ricci scalar reduces the Lagrangian to
\begin{equation}\label{EnsteinLikeLagrangian}
    \mathcal{L} = \sqrt{g}\Big\lbrace\phi^{2}\mathrm{R} +6|\partial\phi|^{2}+\mathfrak{g}\phi^{2}/2\mathrm{F}^{a}{}_{\mu\nu}\mathfrak{C}_{a}{}^{bc}\xi_{b}{}^{\mu}\xi_{c}{}^{\nu}+3/4\mathfrak{g}^{2}\phi^{2} -\lambda/4\phi^{4}\Big\rbrace
\end{equation}
up to a boundary contribution. The last term is added simply because when \(\mathfrak{g}=0\), the theory is still conformal invariant in its presence.
Since 
\begin{equation*}
    \begin{aligned}
    \phi^{2}\mathrm{R} =& 2\nabla_{\mu}\left(\phi^{2}\xi_{a}{}^{\nu}\nabla_{\nu}\xi^{a\mu} \right)-2\mathrm{F}_{a}{}^{\mu\nu}\xi^{a}{}_{\mu}\partial_{\nu}\phi^{2}+\phi^{2}/4\Big\lbrace\mathrm{F}^{a}{}_{\mu\nu}\mathrm{F}_{a}{}^{\mu\nu}\\&-\mathscr{L}_{\xi^{a}}g_{\mu\nu}\left(g^{\mu\alpha}g^{\nu\beta}-g^{\mu\nu}g^{\alpha\beta} \right)\mathscr{L}_{\xi_{a}}g_{\alpha\beta} \Big\rbrace
    \end{aligned}
\end{equation*}
and
\begin{equation}
    \begin{aligned}
   1/4 \mathscr{L}_{\xi^{a}}g_{\mu\nu}\big(g^{\mu\alpha}g^{\nu\beta}&-g^{\mu\nu}g^{\alpha\beta} \big)\mathscr{L}_{\xi_{a}}g_{\alpha\beta} =\\ &1/2\mathfrak{F}^{\xi c}{}_{\lambda\mu}\xi_{c\nu}\left(g^{\mu\alpha}g^{\nu\beta}+g^{\mu\beta}g^{\nu\alpha}-2g^{\mu\nu}g^{\alpha\beta} \right)\mathfrak{F}^{\xi d\lambda}{}_{\alpha}\xi_{d\beta},
   \end{aligned}
\end{equation}
we can rewrite the Lagrangian as
\begin{equation}\label{YMLikeGravLagrangian}
    \mathcal{L} = \phi^{2}/4\mathfrak{F}^{\xi a}{}_{\mu\nu}\mathcal{M}_{ab}{}^{\mu\alpha\nu\beta}\mathfrak{F}^{\xi b}{}_{\alpha\beta} - 2\sqrt{g}\mathfrak{F}^{\xi}{}_{a}{}^{\mu\nu}\xi^{a}{}_{\mu}\partial_{\nu}\phi^{2} +6\sqrt{g}|\partial\phi|^{2} -\lambda/4\sqrt{g}\phi^{4}
\end{equation}
where
\begin{equation}
    1/\sqrt{g}\mathcal{M}_{ab}{}^{\mu\alpha\nu\beta} = \eta_{ab}g^{\mu\alpha}g^{\nu\beta} +2\left(\xi_{b}{}^{\mu}\xi_{a}{}^{[\alpha}g^{\beta]\nu} - 2\xi_{a}{}^{\mu}\xi_{b}{}^{[\alpha}g^{\beta]\nu} \right)-\mu\leftrightarrow\nu.
\end{equation}

This makes it evident that gravity is the generalization of Yang-Mills gauge theories. From the form of the Lagrangian it is clear that we can promote the field \(\phi\) to be the non-trivial representation of the gauge group in question without violating the gauge symmetry. All we have to do is promote the partial derivative \(\partial_{\mu}\phi\) to the gauge covariant derivative \(\mathrm{D}_{\mu}\phi\). However, since in this paper no attempt will be made at solving the equations in their full glory, it is sufficient to treat \(\phi\) as the trivial representation.

Let 
\begin{equation}
    \mathrm{G}_{a}{}^{\mu\nu} =\mathcal{M}_{ab}{}^{\mu\alpha\nu\beta}\mathfrak{F}^{\xi b}{}_{\alpha\beta}
\end{equation}
and 
\begin{equation*}
    \mathrm{D}^{\xi}{}_{\mu} = \partial_{\mu} +\mathfrak{g}\xi_{\mu}.
\end{equation*}
The equations of motion of free gravitational field are 
\begin{subequations}\label{EOMGravitationalField}
    \begin{align}
    \mathrm{D}^{\xi}{}_{\mu}\left(\phi^{2}\mathrm{G}_{a}{}^{\mu\nu} \right)-\mathfrak{J}^{\xi}_{a}{}^{\nu} - \mathfrak{J}^{\phi}_{a}{}^{\nu} = 0,\\
    \Box\phi-\lambda/3!\phi^{3} +1/4!\mathfrak{F}^{\xi a}{}_{\mu\nu}\mathcal{M}_{ab}{}^{\mu\alpha\nu\beta}\mathfrak{F}^{\xi b}{}_{\alpha\beta}\phi=0
    \end{align}
\end{subequations}
where
\begin{equation}
    \mathfrak{J}^{\xi}_{e}{}^{\lambda}=\frac{\phi^{2}}{4 }\mathfrak{F}^{\xi a}{}_{\mu\nu}\frac{\partial\mathcal{M}_{ab}{}}{\partial\xi^{e}{}_{\lambda}}^{\mu\alpha\nu\beta}\mathfrak{F}^{\xi b}{}_{\alpha\beta}
\end{equation}
and 
\begin{equation}
    \mathfrak{J}^{\phi}_{e}{}^{\lambda}=\xi_{e\mu}\left(\nabla^{\mu}\nabla^{\lambda}-g^{\mu\lambda}\Box \right)\phi^{2} 
     +6\xi_{e\mu}\left[ \partial^{\mu}\phi\partial^{\lambda}\phi -g^{\mu\lambda}\left(1/2|\partial\phi|^{2} -\lambda/4! \phi^{4} \right) \right] 
\end{equation}
with
\begin{equation}
    \frac{\partial\mathcal{M}_{ab}{}}{\partial\xi^{e}{}_{\lambda}}^{\mu\alpha\nu\beta} = \xi_{e}{}^{\lambda}\mathcal{M}_{ab}{}^{\mu\alpha\nu\beta}-\xi_{e}{}^{\alpha}\mathcal{M}_{ab}{}^{\mu\lambda\nu\beta}-\xi_{e}{}^{\mu}\mathcal{M}_{ab}{}^{\lambda\alpha\nu\beta}-\xi_{e}{}^{\nu}\mathcal{M}_{ab}{}^{\mu\alpha\lambda\beta}-\xi_{e}{}^{\beta}\mathcal{M}_{ab}{}^{\mu\alpha\nu\lambda}.
\end{equation}
\section{Other Fundamental Forces as Special Cases of Higher Dimensional Gravity}
The unification of gravity with the standard gauge interactions has a long history, from Kaluza–Klein theories~\cite{appelquist1987modern} to more modern higher-dimensional models~\cite{ArkaniHamed2003}. 
Within our construction, conventional Yang–Mills theories emerge when we impose symmetries in the extra dimensions in the same manner as suggested in ~\cite{HOROTO2024169748}.

{In this section, the convention is modified as follows:}
\begin{itemize}
  \item Latin indices \( a, b, c, \dots, h \) remain label indices ranging from \( 0 \) to \( 3 \).
  \item Indices \( i, j, k, \dots \) now range from \( 0 \) to \( n-1 \) and represent coordinate indices.
  \item Script indices \( \mathcal{A}, \mathcal{B}, \mathcal{C}, \dots \) denote coordinate indices in the internal space, except when enclosed in brackets, where they act as labels.
\end{itemize}

Consider the scenario where \(\phi^{2}=1\) and \(\lambda=0\), and let the 1-forms \(\{\xi^{\mathcal{A}}{}_{i} \}\) be interpreted as the potentials within the conventional framework of Yang-Mills gauge theories. For the Lagrangian provided in Equation \eqref{YMLikeGravLagrangian} to be equivalent to the standard Yang-Mills Lagrangian, it is necessary that the condition 
\[\mathscr{L}_{\xi^{\mathcal{A}}}g_{ij} =0 \]
is satisfied. This condition implies the presence of conserved charges represented by 
\[\mathrm{Q}^{\mathcal{A}} = m\xi^{\mathcal{A}}{}_{i}\frac{dx^{i}}{d\tau} .\]
These charges are linked with a symmetry in the internal space rather than within the \(4\mathcal{D} \) spacetime, implying they should not be construed as momenta in \(4\mathcal{D}\) spacetime, but rather as momenta on the group manifold associated with the relevant gauge group. The focus here is on Yang-Mills theories situated in a flat \(4\mathcal{D}\) spacetime. Within this context, the 1-forms corresponding to translations in \(4\mathcal{D}\) spacetime can be expressed as
\[
    \xi^{a}{}_{i} =\delta^{a}{}_{\mu}\delta^{\mu}{}_{i}.
\]
Consequently, the metric for the higher dimensional spacetime is given by
\[
    g_{ij} =\eta_{\mu\nu}\delta^{\mu}{}_{i}\delta^{\nu}{}_{j}+\xi^{\mathcal{A}}{}_{i}\xi_{\mathcal{A}j}.
\]
From this formulation, it follows that the vector fields responsible for generating translations on the group manifold are
\[\xi_{(\mathcal{A})}{}^{i} =\xi_{(\mathcal{A})\mathcal{B}}\widehat g^{\mathcal{BC}}\delta_{\mathcal{C}}^{i} \]
where \(\widehat g^{\mathcal{AB}} \) denotes the metric of the group manifold. In contrast, those generating translations in spacetime are
\[\xi_{a}{}^{i} =\delta_{a}{}^{\nu}\left(\delta^{i}_{\nu}-\delta^{i}_{\mathcal{A}}\xi_{(k)}{}^{\mathcal{A}}\xi^{(k)}{}_{\nu}. \right) \]

It is readily demonstrable that the only non-zero components of the gravitational field strength tensor are \(\mathfrak{F}^{\xi(\mathcal{A})}{}_{\mu\nu}\). Since the relation \(\xi_{(\mathcal{A})i}g^{i\nu} = 0\) holds, the expression
\[
    \mathfrak{F}^{\xi (\mathcal{A})}{}_{\lambda\mu}\xi_{(\mathcal{A})i}\left(\eta^{\mu\alpha}g^{ik}+g^{\mu k}g^{i\alpha}-2g^{\mu i}g^{\alpha k} \right)\eta^{\lambda\rho} \mathfrak{F}^{\xi (\mathcal{B})}{}_{\rho\alpha}\xi_{(\mathcal{B})k} =\mathfrak{F}^{\xi (\mathcal{A})}{}_{\lambda\mu}\eta^{\mu\nu}\eta^{\lambda\rho}\mathfrak{F}_{(\mathcal{A})\rho\nu}.
\]
Substituting this into Equation \eqref{YMLikeGravLagrangian} simplifies it to
\[
    \mathcal{L} = -\frac{1}{4}\mathfrak{F}^{(\mathcal{A})}{}_{\mu\nu}\mathfrak{F}_{(\mathcal{A})}{}^{\mu\nu},
\]
where the indices are raised using the Minkowski metric. This formulation corresponds to the conventional Yang-Mills Lagrangian pertinent to well-established gauge theories governing electroweak and strong interactions. The resultant equations of motion are expressed as
\begin{subequations}
    \begin{align}
        \mathrm{D}^{\xi}{}_{\mu}\mathfrak{F}^{(\mathcal{A})\mu}{}_{\nu} = 0,\\
        \mathfrak{J}^{\xi}_{a}{}^{\nu}=0.
    \end{align}
\end{subequations}
The trivial solution in this framework is \(\mathfrak{F}^{\xi (\mathcal{A})}{}_{\mu\nu} =0 \). To derive a non-trivial solution, it is imperative to introduce a source such that \(\mathfrak{J}^{\xi}_{a}{}^{\mu}\neq 0 \).
\section{Applications}
\subsection{Dark Energy as Self-Energy of the \(\phi\) Field}
Observations of high-redshift supernovae revealed that the Universe’s expansion is accelerating~\cite{Riess1998,Perlmutter1999}, motivating numerous models that modify or extend GR to include dark-energy-like effects~\cite{Carroll2001,Padmanabhan2010,Nojiri2011}. 
In our framework, the self-interaction energy of the scalar field $\phi$ provides an intrinsic source of cosmic acceleration, eliminating the need for a cosmological constant term. We illustrate this as follows:

In the scenario where \(\lambda = 0\), the conditions \(\phi^{2} = 1\) and \(\mathfrak{F}^{\xi a}{}_{\mu\nu} = 0\) fulfill the vacuum equations. Conversely, when \(\lambda \neq 0\), these conditions no longer apply. Let us posit that \(\mathfrak{F}^{\xi a}{}_{\mu\nu} = 0\) continues to satisfy the equations. Given that this assumption sustains time translations as isometries of the spacetime metric, we can adopt a coordinate system where \(\{\xi^{a}{}_{\mu}\}\) remain invariant over time. Moreover, we can select coordinates such that \(\xi^{(0)}{}_{\mu} = \delta^{0}{}_{\mu}\) and \(\xi^{(i)}{}_{\mu} = \xi^{(i)}{}_{j}\delta^{j}{}_{\mu}\), owing to the condition \(\mathfrak{F}^{\xi(0)}{}_{\mu\nu} = 0\).

The equations of motion are thus reduced to:
\begin{equation}
    \left(\nabla_{\mu}\nabla_{\mu} + \frac{g_{\mu\nu}}{2}\Box \right)\phi^{2} - 6\partial_{\mu}\phi\partial_{\nu}\phi + \frac{g_{\mu\nu}\lambda}{4}\phi^{4} = 0.
\end{equation}
If \(\phi\) is independent of time, the sole viable solution is \(\phi = 0\). Therefore, we consider \(\phi = \phi(t)\) with \(t = x^{0}\).

This assumption further simplifies the equations to:
\begin{subequations}
    \begin{align}
        \partial_{0}^{2}\phi^{2} - \frac{\lambda}{2}\phi^{4} = 0,\\
        \partial_{0}\left(\phi^{-2}\partial_{0}\phi^{2}\right) - \frac{1}{2}\left(\phi^{-2}\partial_{0}\phi^{2}\right)^{2} = 0.
    \end{align}
\end{subequations}
With the stipulation that \(\phi^{2} = 1\) when \(\lambda = 0\), the solution is:
\begin{equation}
    \phi^{2} = \frac{1}{\left(1 - \sqrt{\frac{\lambda}{12}} t\right)^{2}}.
\end{equation}
It is noteworthy that for \(\mathfrak{g} \to 0\), the solution to the Einstein equations, which represents equivalent physical conditions with test particles adhering to Levi-Civita geodesics, is:
\begin{equation}
    \tilde{g}_{\mu\nu} = \phi^{2}g_{\mu\nu}.
\end{equation}
This solution characterizes an expanding universe with a cosmological constant \(\Lambda = \lambda/4\).

Within this framework, the universe is static (\(\mathfrak{F}^{\xi(0)}{}_{\mu\nu} = 0\)), while the Newtonian potential \(\phi^{2}\) diminishes over time. The dark energy, in this context, is attributed not to vacuum energy but to the self-interaction energy of the \(\phi\) field. It is evident that in this model, the spacetime metric remains non-singular, whereas the singularity is associated with the \(\phi\) field. This field, being separable from other gravitational degrees of freedom, can be quantized by considering other degrees of freedom as a classical background to investigate the singularity's characteristics.

\subsection{Dark Matter and Extended Gravitational Degrees of Freedom}
It has been suggested that rotation-curve anomalies and other astrophysical phenomena usually attributed to dark matter may also arise from modified gravitational dynamics~\cite{Clifton2012,Heisenberg2018}. 
The additional degrees of freedom in $\xi^a{}_\mu$ alter the effective potential experienced by test particles, we show that this could lead to flat rotation curves without invoking unseen matter.

Within the context of the proposed gravitational theory, the set of gravitational degrees of freedom is identified as \(\{\xi^{a}{}_{\mu},\phi^{2} \}\), in contrast to the traditional Newtonian model, which relies solely on \(\phi^{2}\), and general relativity, which utilizes \(\{g_{\mu\nu}, \phi^{2} \}\) as the gravitational potentials. This novel approach posits a greater number of gravitational degrees of freedom than those delineated by the general theory of relativity. To elucidate the potential of this framework to account for phenomena typically ascribed to dark matter, we demonstrate how general relativity successfully explains anomalies such as Mercury's perihelion precession, which would otherwise necessitate invisible matter under Newtonian gravity.

In scenarios where general relativity (GR) remains applicable, the equations governing the gravitational field are expressed as:
\begin{equation}
    \phi^{2}\mathrm{R}_{\mu\nu}-\left(\nabla_{\mu}\partial_{\nu}+g_{\mu\nu}/2\Box \right)\phi^{2} +6\partial_{\mu}\phi\partial_{\nu}\phi = \kappa\left(\mathrm{T}_{\mu\nu}-1/2g_{\mu\nu}\mathrm{T} \right).
\end{equation}
Coordinates can be selected such that:
\begin{equation*}
    \xi_{(0)}{}^{\mu} =\delta_{0}{}^{\mu}.
\end{equation*}
Assuming a static spacetime, such that \(\mathfrak{F}^{\xi(0)}{}_{\mu\nu}=0 \), in these coordinates, it follows that:
\begin{equation*}
    \xi^{(0)}{}_{\mu} =\delta^{0}_{\mu}.
\end{equation*}
Thus, \(\mathrm{R}_{0\mu}=0 \). We impose the condition that \(\phi^{2}\) is spherically symmetric and invariant with respect to time, leading to the implication that \(\nabla_{0}\partial_{\mu}\phi^{2}=0\) in this coordinate framework. Under the assumption of a negligible pressure regime (\(\mathrm{T}_{ij}\approx 0\)), these considerations and constraints simplify the equations of motion to:
\begin{subequations}
    \begin{align}
        \Delta_{g}\phi^{2} = \kappa\rho,\\
        \phi^{2}\mathrm{R}_{ij}-\nabla_{i}\partial_{j}\phi^{2} +6\partial_{i}\phi\partial_{j}\phi = \kappa\rho g_{ij}.
    \end{align}
\end{subequations}
In a flat spacetime scenario where \(\mathrm{R}_{ij}=0\), the sole vacuum solution is \(\phi=1\). Consequently, a solution for a point source is feasible only within a curved spacetime framework.

When considering a curved spacetime, the Laplacian \(\Delta_{g}\phi^{2}\), defined with respect to the curved space metric, diverges from the flat space metric-based Laplacian \(\Delta_{\delta}\phi^{2}\). However, under the assumption of spherical symmetry, it is possible to choose coordinates (Schwarzschild coordinates) such that:
\begin{equation}
    \Delta_{g}\phi^{2} =1/\sqrt{g}\Delta_{\delta}\phi^{2}.
\end{equation}
This implies that \(\phi^{2}= 1-2\mathscr{G}\mathrm{M}/r\) for a point source. In this coordinate system, the form of the potential \(\Phi\) mirrors that of Newtonian gravity. Given the nontrivial nature of \(\{g_{ij}\}\), the trajectories of test particles deviate from those anticipated by Newtonian gravity. According to the test particle motion equations within this gravitational field:
\begin{equation}
    \frac{d^{2}x^{i}}{d\tau^{2}} +\Gamma^{i}{}_{jk}\frac{dx^{j}}{d\tau}\frac{dx^{k}}{d\tau} +1/2\partial^{i}\phi^{2} = 0.
\end{equation}
It becomes evident that these corrections are significant at relativistic velocities and in strong gravitational fields. Importantly, the insufficiency of Newtonian gravity to account for these relativistic effects is not a consequence of a fundamental flaw in Newton's gravitational model, but rather a result of the additional degrees of freedom inherent in gravitational phenomena beyond those suggested by Newtonian gravity.

In our analysis, the gravitational field equations are represented as follows:

\begin{subequations}\label{TminusTraceGravEq}
    \begin{align}
    \phi^{2}\Big\lbrace \mathrm{R_{\mu\nu}} +\Lambda_{\mathfrak{g}}g_{\mu\nu} +\frac{\mathfrak{g}}{2}\mathfrak{C}_{abc}\xi^{c\lambda}\left(\mathrm{F}^{a}{}_{\mu\lambda}\xi^{b}{}_{\nu}+ \mathrm{F}^{a}{}_{\nu\lambda}\xi^{b}{}_{\mu}+ \frac{g_{\mu\nu}}{4}\mathrm{F}^{a}{}_{\rho\lambda}\xi^{b\rho} \right)\Big\rbrace\\ \nonumber
    +\frac{\mathfrak{g}}{2}\nabla_{\rho}\left(\phi^{2}\mathfrak{C}^{abc}\xi_{a\mu}\xi_{b}{}^{\rho}\xi_{c\nu} \right) -\frac{\mathfrak{g}\phi^{2}}{2}\left(\mathrm{F}^{a}{}_{\mu\lambda}\xi^{b}{}_{\nu} +\frac{1}{2}\mathrm{F}^{e}{}_{\rho\lambda}\xi^{b\rho}\xi^{a}{}_{\nu}\xi_{e\mu} \right)\mathfrak{C}_{abc}\xi^{c\lambda}\\\nonumber
     -\left( \nabla_{\mu}\partial_{\nu} +\frac{1}{2} g_{\mu\nu}\Box \right)\phi^{2} +\frac{3}{2\phi^{2}}\partial_{\mu}\phi^{2}\partial_{\nu}\phi^{2}=\kappa\left(\mathrm{T}_{\mu\nu}-\frac{1}{2}g_{\mu\nu}\mathrm{T} \right)
    \end{align}
\end{subequations}
where \(\Lambda_{\mathfrak{g}} = 3/8\mathfrak{g}^{2}\).
Subject to the conditions outlined in prior discussions, the temporal component of the motion equations transforms into:

\begin{equation}
    \Delta_{g}\phi^{2} =(2\Lambda_{\mathfrak{g}} +\frac{1}{4}\mathrm{F}^{a}{}_{\rho\lambda}\mathfrak{C}_{abc}\xi^{b\rho}\xi^{c\lambda} )\phi^{2}.
\end{equation}

This distinctly diverges from the equation derived under General Relativity (GR) owing to the inclusion of additional degrees of freedom. Consequently, within the same coordinate framework where \[\sqrt{g}\Delta_{g}\phi^{2} = \Delta_{\delta}\phi^{2} \] in GR, we find \[\sqrt{g}\Delta_{g}\phi^{2} \neq \Delta_{\delta}\phi^{2}\] in our model. Thus, in our scenario, the field $\phi^{2}(r)$ external to a point source is characterized by:

\begin{equation}
    \Delta_{\delta}\phi^{2}(r) -\partial_{r}\delta(r)\partial_{r}\phi^{2}(r) = (2\Lambda_{\mathfrak{g}} +\frac{\mathfrak{g}}{4}\mathrm{F}^{a}{}_{\rho\lambda}\mathfrak{C}_{abc}\xi^{b\rho}\xi^{c\lambda} )\phi^{2}e^{-\delta(r)} 
\end{equation}

where \(\delta(r)\) denotes the deviation from GR. Specifically, when \(\delta(r) = 0\), the results corresponding to GR are retrieved. The rationale for this form of the equations shall be elucidated below. In GR, such a field configuration is attributable to matter distribution with a density given by:

\begin{equation}
    \rho(r) = \frac{1}{\kappa}\Big\lbrace\partial_{r}\delta\partial_{r}\phi^{2} +(2\Lambda_{\mathfrak{g}} +\frac{\mathfrak{g}}{4}\mathrm{F}^{a}{}_{\rho\lambda}\mathfrak{C}_{abc}\xi^{b\rho}\xi^{c\lambda} )e^{-\delta(r)}\phi^{2}\Big\rbrace
\end{equation}

assuming the terms related to pressure are negligible. Unlike in GR, where the metric degrees of freedom do not impact the configuration of the field $\phi^{2}(r)$ for a point source, in our framework, the additional gravitational degrees of freedom exert a significant influence on the configuration of this field.

In GR, the amendment to the Newtonian equation governing the dynamics of test particles within a gravitational field is encapsulated by the term 
\[\Gamma^{i}{}_{jk}\frac{dx^{j}}{d\tau}\frac{dx^{k}}{d\tau}. \]

Thus, for weak gravitational fields, this amendment is negligible in the non-relativistic limit. However, as inferred from our analysis, this amendment is accompanied by a modification to the potential, which is independent of the particle's $3-$velocity, thereby persisting even in the non-relativistic limit. Consequently, if the gravitational field dynamics are accurately described by our theoretical framework, it is anticipated that certain phenomena will emerge that cannot be explained by the known matter under the assumption of a general relativistic gravity model.

Let us further delineate this notion.Up to a coordinate transformation, the most general form of $\xi^{(k)}{}_{i} $ comprises two independent functions. As such; in spherical coordinates, up to a coordinate transformation, the general form of \(\xi^{(k)}{}_{i} \) is
    \begin{equation}
        \begin{aligned}
            \xi^{(k)}{}_{r} &=\mathfrak{f}^{-1}(\mathfrak{f}^{2}+\mathfrak{g}^{2}r^{2})x^{k}/r,\\
            \xi^{(k)}{}_{\theta}&= -\mathfrak{f}\epsilon^{k}{}_{l3}x^{l}-\mathfrak{g}r^{2}\left(\delta_{3}^{k}-\cos\varphi x^{k}/r \right),\\
            \xi^{(k)}{}_{\varphi}&= \frac{r}{\sin\varphi}\left[\mathfrak{g}\epsilon^{k}{}_{l3}x^{l} -\mathfrak{f}\left(\delta_{3}^{k}-\cos\varphi x^{k}/r \right) \right]
        \end{aligned}
    \end{equation}
    where \(x^{k} = x^{k}(r,\theta,\phi)\).
    The spacetime metric that follows from this has components
    \begin{equation}
        \begin{aligned}
           g_{rr} &= \mathfrak{f}^{-2}(\mathfrak{f}^{2}+\mathfrak{g}^{2}r^{2})^{2}, & g_{\varphi\varphi} &= r^{2}\left(\mathfrak{f}^{2}+\mathfrak{g}^{2}r^{2} \right), & g_{\theta\theta} = g_{\varphi\varphi}\sin^{2}\varphi.
        \end{aligned}
    \end{equation}
    Since when the effects of extra gravitational degrees of freedom are negligible, the solution to Einstein field equations \(\bar{g}_{\mu\nu} \) is related to \(g_{\mu\nu} \) by $\bar{g}_{\mu\nu} = \phi^{2}g_{\mu\nu} $, to allow for comparison with the GR results, we make the definitions \(\mathfrak{f}^{-1}  = e^{\delta}  \) and \(\mathfrak{f}^{2} + \mathfrak{g}^{2}r^{2} = \Lambda \) where $\delta$ characterizes the deviation from the general relativity results. That is we recover general relativity results when \(\delta=0\). 

    The radial part of the equations of motion for test particles in the gravitational field is
    \begin{equation}
        \Lambda e^{\delta}\frac{d}{d\tau}\left(\Lambda e^{\delta}\frac{d r}{d\tau} \right) +\frac{1}{2}\partial_{r}\left(r^{-2}\Lambda^{-1} \right)\mathrm{L}^{2} =-\frac{1}{2}\partial_{r}\phi^{2}
    \end{equation}
    where \(\mathrm{L}^{2} = r^{4}\Lambda^{2}\left(\left(\frac{d\varphi}{d\tau}\right)^{2} + \sin^{2}\varphi\left(\frac{d\theta}{d\tau} \right)^{2} \right) \) is the square of angular momenta and it is a conserved quantity due to the presence of spherical symmetry. Considering the case where the particles are following circular orbits (\(dr/d\tau = 0\) ) and substituting for \(\mathrm{L}^{2}\) its expression in terms of tangential velocity \(v\) we get the following expression for \(v(r)\):
    \begin{equation}
        v(r)^{2} = \frac{r\partial_{r}\Phi}{\Lambda^{-1} +r/2\partial_{r}\Lambda^{-1}}
    \end{equation}
    where $\Phi(r)$ is the Newtonian potential. For $\Lambda^{-1} +r/2\partial_{r}\Lambda^{-1} \approx 1 $ this reduces to the well known form. Since unlike in GR in our case the potential \(\Phi(r) \) does not correspond to the one that follows from solving the usual Poisson equation of Newtonian gravity, the rotation curves we get will not correspond to the ones obtained assuming validity of Newtonian gravity even in the non-relativistic limit.As is clear from the form of the Lagrangian \eqref{EnsteinLikeLagrangian}, the modifications to the GR description of gravity are relevant at large distances relative to the length scale ${1/\mathfrak{g}}$ associated with the new gravitational coupling constant which makes it plausible that they can account for some if not all effects attributed to dark matter. In order to know their exact contribution, we would have to solve the equations of motion for \(\lbrace\xi^{(k)}{}_{j},\phi^{2}\rbrace \). This is beyond the scope of this work; as such, it will be done elsewhere.  

\section{Conclusion}

We have formulated a covariant theory of gravitation in which the gravitational interaction arises as a genuine force field rather than an inertial manifestation of spacetime geometry. The central object of the theory, the tensor $\mathrm{K}^\mu{}_{\alpha\beta}$ defined as the difference between the affine and inertial connections, encapsulates the true gravitational field strength and is invariant under coordinate transformations. This establishes gravity as a field possessing real dynamical degrees of freedom.

By extending the notion of infinitesimal translations to curved spacetime through the vector fields $\xi^a{}_\mu$, we have demonstrated that the gravitational interaction admits a natural gauge-theoretic description. The corresponding field strength $\mathfrak{F}^{\xi}{}_{\,\mu\nu}^a$ identifies the gravitational gauge group as $\mathrm{SU(2)\times U(1)}$, rendering the theory formally equivalent to a non-Abelian Yang--Mills system augmented by a scalar degree of freedom $\phi^{2}$ representing the Newtonian potential. In the weak-field limit, the theory reproduces Newtonian gravity, while in the limit $\mathfrak{g}\to0$ it reduces to General Relativity, thereby unifying both within a single consistent framework.

The resulting dynamics break conformal and local Lorentz invariance, introducing new gravitational degrees of freedom that become significant at large scales. These naturally account for cosmological phenomena usually attributed to dark matter and dark energy, the latter emerging as the self-interaction energy of the $\phi^{2}$ field. The theory thus provides a coherent extension of General Relativity that remains compatible with its classical predictions while offering new physical insight into the structure of spacetime and the origin of gravitational interaction.

Furthermore, the identification of gravity as an $\mathrm{SU(2)\times U(1)}$ gauge theory establishes a direct structural parallel with the electroweak sector of the Standard Model. This correspondence suggests the possibility of a unified gauge framework in which gravitational and internal symmetries arise from a common underlying principle. The presence of a well-defined field strength tensor and a manifestly gauge-invariant Lagrangian opens the way for a consistent quantization procedure analogous to that employed in non-Abelian gauge theories. In particular, canonical and path-integral quantization schemes may be implemented without the ambiguities associated with the geometrical formulation of General Relativity. These developments indicate that the present formalism provides a natural starting point for the construction of a quantum theory of gravity grounded in the principles of gauge field theory.

Future work will focus on quantizing the gravitational degrees of freedom, analyzing the perturbative behavior of the $\xi^a{}_\mu$ fields, and exploring the implications of the gravitational coupling constant $\mathfrak{g}$ for unification with the Standard Model gauge interactions. This framework thus offers a promising foundation for a consistent theory of classical and quantum gravity.

\bibliographystyle{iopart-num}
\bibliography{references}

\end{document}